\documentclass[journal]{IEEEtran}
\IEEEoverridecommandlockouts
\usepackage{cite}
\usepackage{amsmath,amssymb,amsfonts}
\usepackage{algorithmic}
\usepackage{graphicx}
\usepackage{textcomp}
\usepackage{xcolor}
\usepackage{url}
\usepackage{float}
\usepackage{soul}
\usepackage{stackengine}

\usepackage{amsthm}
\newtheorem{theorem}{Theorem}

\newtheorem{remark}{Remark}

\newtheorem{assumption}{Assumption}

\usepackage{mathrsfs}
\DeclareMathOperator*{\argmin}{arg\,min}

\usepackage[most]{tcolorbox}
\usepackage{color}
\usepackage{xcolor,colortbl}
\definecolor{Gray}{gray}{0.85}

\begin{document}

\title{Breaking the Barriers of One-to-One Usage of~Implicit Neural Representation in Image~Compression: A Linear Combination Approach with Performance Guarantees}

\author{\IEEEauthorblockN{Sai Sanjeet,~\IEEEmembership{Student Member,~IEEE}, Seyyedali Hosseinalipour,~\IEEEmembership{Member,~IEEE}, Jinjun Xiong,~\IEEEmembership{Senior Member,~IEEE}, Masahiro Fujita,~\IEEEmembership{Life Member,~IEEE}, and Bibhu Datta Sahoo,~\IEEEmembership{Senior Member,~IEEE}}\\
\thanks{Sai Sanjeet, Seyyedali Hosseinalipour, and Bibhu Datta Sahoo are with the Department of Electrical Engineering, University at Buffalo--SUNY, Buffalo, NY, US (e-mail: \{syerragu, alipour, bibhu\}@buffalo.edu).}
\thanks{Jinjun Xiong is with the Department of Computer Science and Engineering, University at Buffalo--SUNY, Buffalo, NY, US (e-mail: jinjun@buffalo.edu).}
\thanks{Masahiro Fujita is with System Design Research Center, University of Tokyo, Tokyo, Japan (e-mail: fujita@ee.t.u-tokyo.ac.jp).}
\vspace{-4.5mm}
}

\maketitle

\begin{abstract}
In an era where the exponential growth of image data driven by the Internet of Things (IoT) is outpacing traditional storage solutions, this work explores and advances the potential of Implicit Neural Representation (INR) as a transformative approach to image compression. INR leverages the function approximation capabilities of neural networks to represent various types of data. While previous research has employed INR to achieve compression by training small networks to reconstruct large images, this work proposes a novel advancement: representing multiple images with a single network. By modifying the loss function during training, the proposed approach allows a small number of weights to represent a large number of images, even those significantly different from each other. A thorough analytical study of the convergence of this new training method is also carried out, establishing upper bounds that not only confirm the method's validity but also offer insights into optimal hyperparameter design. The proposed method is evaluated on the Kodak, ImageNet, and CIFAR-10 datasets. Experimental results demonstrate that all 24 images in the Kodak dataset can be represented by linear combinations of two sets of weights, achieving a peak signal-to-noise ratio (PSNR) of 26.5 dB with as low as 0.2 bits per pixel (BPP). The proposed method matches the rate-distortion performance of state-of-the-art image codecs, such as BPG, on the CIFAR-10 dataset. Additionally, the proposed method maintains the fundamental properties of INR, such as arbitrary resolution reconstruction of images.
\end{abstract}

\begin{IEEEkeywords}
Implicit Neural Representation (INR), image compression, bits per pixel (BPP), constrained training, Kodak dataset, Imagenet, CIFAR10.
\end{IEEEkeywords}

\vspace{-2.5mm}
\section{Introduction}

In the digital age, the proliferation of image data driven by the Internet of Things (IoT) has resulted in an unprecedented increase in the volume of images generated and stored. This surge has prompted numerous studies focused on managing and compressing this data \cite{Wang2021, Chakraborty2021, Wang2022, Chamain2022, An2024}. Innovative approaches to image compression are essential to address the challenges posed by this rapid data expansion. Implicit Neural Representation (INR) has emerged as a promising technique for image compression, yet it faces fundamental limitations. This work seeks to overcome one such limitation: the need to use a separate network for each image.

Implicit Neural Representation (INR) is a new class of neural networks that can be used to learn complex representations of data \cite{Park2019,Mescheder2019,Chen2019,Sitzmann2020} by exploiting the function approximation property of neural networks \cite{Michalkiewicz2019,Morreale2022}. The goal of INR is to learn a function that maps a point in the input space to a point in the output space. Recent breakthroughs have demonstrated the effectiveness of INR in learning continuous representations of various functions, including but not limited to images, videos, 3D models, and radiance fields \cite{Chen2021,Chen2022,Luigi2023,Mildenhall2020}. The learned representations can be used for various tasks such as image generation, super-resolution, and shape reconstruction.

Neural networks were shown to be function approximators back in the 1980s \cite{Hornik1989}. The universal approximation theorem states that a feedforward neural network with a single hidden layer can approximate any continuous function on a compact subset of $n$-dimensional space $\mathbb{R}^n$ \cite{Cybenko1989}. INRs serve as prime examples of how this fundamental theoretical principle can be exploited in practical scenarios.
In their basic form, INR models are fully connected networks that take in independent variables and predict the functional values. The weights/parameters of the trained network implicitly represent the function at hand. The choice of the network size and the activation function used in the network determine the nature and complexity of the function that can be represented. The work of Sitzmann \textit{et al.} \cite{Sitzmann2020} showed that INR models with periodic activation functions, such as the sine activation function, can represent complex real-world functions. Their proposed model, called the SIREN model, was shown to be particularly effective in representing images.

\begin{figure}[htbp]
    \centering
    \vspace{-1em}
    \includegraphics[scale=0.8]{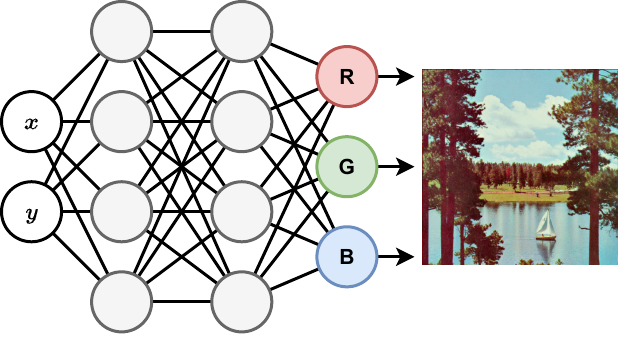}
    \vspace{-1em}
    \caption{A schematic of an INR model used for image representation.} \vspace{-0.5em}
    \label{fig:INR}
\end{figure}

Fig. ~\ref{fig:INR} depicts a schematic of an INR model used for image representation/reconstruction. The depicted INR model is a fully connected network with two inputs $x$ and $y$ (i.e., the pixel coordinates) and three outputs (i.e., the predicted RGB values of the input pixel). In the conventional INR training, the model is specifically trained to overfit on a single image. The inputs and the outputs are typically normalized to the range $[-1, 1]$ \cite{Sitzmann2020}. The loss function used for training is often the mean squared error between the predicted RGB values and the respective ground truth values. After the training period concludes, the overfit model captures the parameters necessary to represent the input image accurately. These parameters can then be utilized to regenerate or reconstruct the image. This capability is particularly advantageous when dealing with high-resolution images, where an INR network, despite having a relatively small number of parameters, can be trained to regenerate the entire image effectively. This feature facilitates image compression. Specifically, rather than transmitting a high-resolution image across a networked system (e.g., an edge-to-cloud wireless platform), a compact  INR network with few model parameters can be transferred instead. This approach can significantly reduce transmission overhead (e.g., in terms of energy consumption and latency).

Image compression using INR has recently gained a lot of attention leading to several works in the field \cite{Dupont2021,Dupont2022,Strumpler2022,Ladune2023,Pham2023,Lee2023,Girish2023}. Initial works of Dupont \textit{et al.} (COIN) \cite{Dupont2021} showed the effectiveness of using INR where a simple SIREN network trained on a single image outperformed the JPEG compression standard in terms of peak signal-to-noise ratio (PSNR) at low bit rates. The work of Strumpler \textit{et al.} \cite{Strumpler2022} extended the idea by meta-learning the initialization of weights and using entropy coding to gain further improvements in compression performance. Ladune \textit{et al.} \cite{Ladune2023} proposed a method where a hierarchical latent representation of images is learned using INR (COOL-CHIC). The method showed competitive performance that is on par with the state-of-the-art image codecs.

Most of the recent works have focused on improving the performance of INR-based image compression by utilizing predetermined or meta-learning encodings along with entropy coding for further compression \cite{Pham2023,Lee2023,Girish2023}. However, to the best of our knowledge, no work has explored beyond the fundamental limitation of INR: \textit{representing one image with one network}. This work proposes a novel advancement in the field by addressing this fundamental limitation, opening a new avenue of research focusing on breaking the barriers on one-to-one usage of INR networks in image compression and reconstruction. In particular, our proposed approach enables the training of a small number of INR networks, which can be used in conjunction to represent a large number of images. This is achieved by modifying the loss function used for training, allowing a linear combination of INR weights to represent multiple images, even if the images are significantly different from each other. We further complement our methodology with a set of analytical convergence analysis, which reveal the soundness of our method and provide guidelines on the choice of hyperparameters, the details of which are discussed in Section \ref{sec:convergence}. Experimental results show that the proposed method can represent as many as $128$, $24$, and $96$ images with only {\bf two} INR networks for CIFAR-10, Kodak, and ImageNet datasets, respectively, with PSNR $\ge 18$ dB. For the CIFAR-10 dataset, the proposed method outperforms prior INR-based methods and closes the gap with BPG, a state-of-the-art image codec.


This manuscript is organized as follows. Section \ref{sec:proposed_training} describes the motivation and formulation of the proposed training method. Section \ref{sec:convergence} provides a detailed analysis of the convergence bounds of the proposed training method. Section \ref{sec:benchmarking} presents the experimental results on the Kodak, ImageNet, and CIFAR-10 datasets, and compares the proposed method with prior works. Finally, Section \ref{sec:conclusion} concludes the manuscript.

\vspace{-2mm}
\section{Proposed Training Method}
\label{sec:proposed_training}
\begin{figure}[htbp]
    \centering
    \includegraphics[scale=0.8]{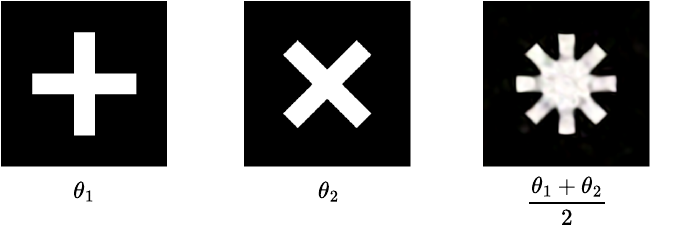}
    \vspace{-2.65em} 
    \caption{Averaging the weights of two INR models results in an approximate average of the two images.} \vspace{-.1em}
    \label{fig:average}
\end{figure}

The proposed training method is inspired by the following observation. Two INR models were trained independently on two different images, with the same architecture, hyperparameters, and initial values of weights. When trying to combine the weights of the two models with operations such as element-wise average, it was found that the combined weights represent an approximate average of the two original images. In the visual example depicted in Fig.~\ref{fig:average}, one model is trained on the `$+$' sign to obtain $\theta_1$ and another model is trained on the `$\times$' sign to obtain $\theta_2$. The weights of the two models are combined to obtain $\theta_{\text{avg}} = (\theta_1 + \theta_2)/2$. The weights of the combined model are used to generate the image. The generated image is approximately an `$*$' sign, which is the pixel-wise average of the two original images.

\begin{figure}[htbp]
    \centering
    \includegraphics[scale=0.81]{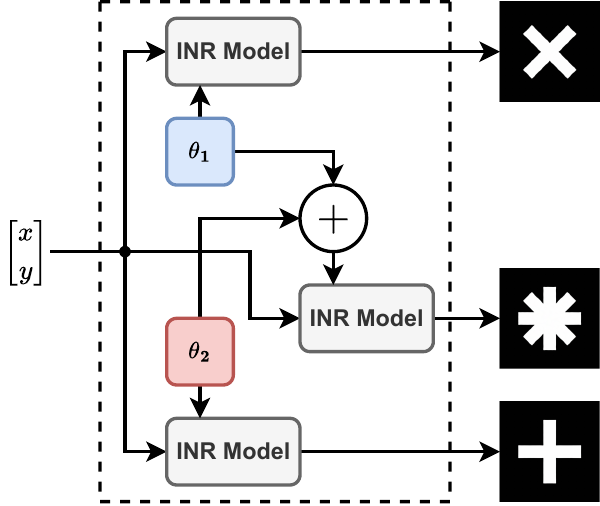} \vspace{-1em}
    \caption{Proposed method of representing the average of two images using the weights of two INR models.}
    \label{fig:training_model}
\end{figure}

This observation suggests that the weights of the INR models, each trained on a single image, can be combined to represent a new image that is a linear combination of original images. To verify if the linear combination can perfectly generate the average image, we further modify the training procedure to include the loss function for the average image. Fig. \ref{fig:training_model} shows the structure of our modified training method.

\begin{figure}[htbp]
\vspace{-0mm}
    \centering
    \includegraphics[scale=0.81]{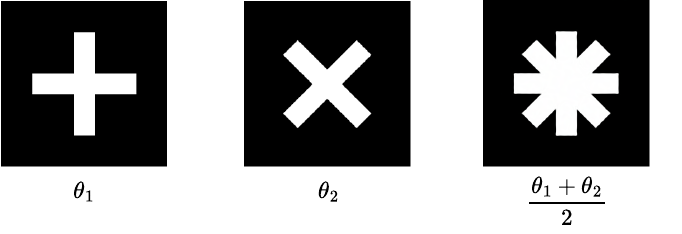}
     \vspace{-8mm} 
    \caption{Averaging the weights of two INR models trained in a constrained manner results in the exact average of the two images. Note that $\theta_1$ and $\theta_2$ here are different from the $\theta_1$ and $\theta_2$ in Fig. \ref{fig:average}. For simplicity, we use the same notation without ambiguity.} \vspace{-2mm}
    \label{fig:average_perfect}
\end{figure}

In Fig. \ref{fig:training_model}, $\theta_1$ and $\theta_2$ are the trainable weights that are trained to generate the `$+$' and `$\times$' signs, respectively. The weights are combined to generate the average image. The loss function is the mean squared error between all the three predicted and ground truth images. With this training method, $\theta_1$ and $\theta_2$ were found to be such that they not only can produce the `$+$' and `$\times$' signs, but also their average can perfectly generate the `$*$' sign, as shown in Fig.~\ref{fig:average_perfect}. This is reasonable as, for example, the gradients computed during training for updating $\theta_1$ come from the loss function of both the primary `$+$' and secondary `$*$' signs.

\begin{figure}[htbp]
    \centering
    \vspace{-2mm}
    \includegraphics[scale=0.8]{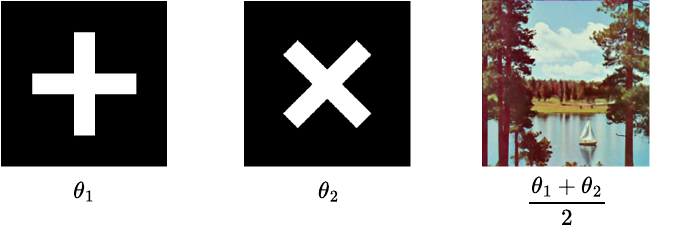}
     \vspace{-8mm} 
    \caption{Averaging the weights of two INR models trained in a constrained manner can represent a completely different image.} \vspace{-0.5em}
    \label{fig:average_diff}
\end{figure}

The more aggressive question was the following. Can the weights of the INR model be combined to represent an image that is completely different from the original images? Using the same training method, we find that this is indeed feasible, as shown in Fig.~\ref{fig:average_diff}. In the illustrative example in Fig.~\ref{fig:average_diff}, we reveal that the weights of the two INR models trained on black and white images of `$+$' and `$\times$' signs can be trained in such a way that their combinations can generate the colored sailboat image. This powerful observation indicates that the weights of the INR model can be combined to represent an image that is completely different from the original images. 

This work exploits these observations and proposes a novel training method for INR models that allows the representation of a large number of images using linear combinations of a few sets of INR network weights.  Let $Y = F(X;\theta)$ be the forward pass of an INR model with weights $\theta$, input $X$, and output $Y$, and  $\widehat{Y}$ be the ground truth image. The aim of conventional INR training is to find the optimal weights $\theta^\star$ that minimize the loss function $\mathcal{L}(\theta) = \mathscr{E}(Y, \widehat{Y})$, {\em i.e.}, $\theta^\star \triangleq \argmin_{\theta} \mathcal{L}(\theta)$, where $\mathscr{E}(.,.)$ is a metric of distance (e.g., $\ell_2$ norm). In this work, we aim to extend this conventional formulation, which considers training one INR network for one image, to a scenario in which $N$ INR network weights are used conjunctly to represent $M$ images, where $N \leq M$.

To formulate the learning goal of our interest, we first consider INR training via considering $N$ sets of trainable weights $\Theta = [\theta_1, \theta_2, \ldots, \theta_N]$ and $M$ images $\widehat{Y}_1, \widehat{Y}_2, \ldots, \widehat{Y}_M$. We then modify the loss function to capture the error in representing all the images using the $N$ sets of weights. The modified loss function is given by (\ref{eq:loss})
 and (\ref{eq:loss_i}). In practice, the error function in (\ref{eq:loss_i}) is the mean squared error between the predicted and the ground truth images, or any other appropriate metric \vspace{-0.5em}
\begin{align}
    \label{eq:loss}
    \mathcal{L}(\Theta) &= \sum_{i=1}^{M} \gamma_i \mathcal{L}_i(w_i), \\
    \mathcal{L}_i(w_i) &= \mathscr{E}(F(X, w_i), \widehat{Y}_i) ,~~~ 1\leq i\leq M,
    \label{eq:loss_i}
\end{align}

\noindent where $\gamma_i$ are predetermined constants , such that $\gamma_i\in (0,1)$, $\forall i$, and $\sum_{i=1}^M\gamma_i=1$, and $w_i$ are linear combinations of the $N$ sets of INR network weights, given by \vspace{-0.5em}
\begin{equation}
    w_i = \sum_{j=1}^{N} \alpha_{ij} \theta_j,~~~ 1\leq i\leq M,
    \label{eq:weights}
\end{equation}
\noindent where $\alpha_{ij}$ are the weighting coefficients. It is to be noted that $\Theta$ are the only set of weights that are updated during training. In particular, in our paradigm, the goal of INR training is obtaining the set $\Theta^\star$, where $\Theta^\star \triangleq \argmin_{\Theta}\mathcal{L}(\Theta)= \sum_{i=1}^{M} \gamma_i \mathcal{L}_i(w_i)$. Let the gradients of the error function with respect to $w_i$ be given by $\nabla_{w_i} \mathcal{L}_i(w_i)$. Since $w_i$ appears in only one term of the summation in (\ref{eq:loss}), the gradient of the loss function with respect to $w_i$'s is $\nabla_{w_i} \mathcal{L}(\Theta) = \gamma_i \nabla_{w_i} \mathcal{L}_i(w_i)$. Using the chain rule of differentiation, the gradient of the loss function with respect to $\theta_j$'s can be computed as follows:\footnote{The weights $\theta_j$'s are linearly combined with $\alpha_{ij}$'s to obtain $w_i$'s, and the loss functions are linearly combined with $\gamma_i$'s to obtain the total loss.}\vspace{-0.5em}
\begin{equation}
    \nabla_{\theta_j} \mathcal{L}(\Theta) = \sum_{i=1}^{M} \alpha_{ij} \gamma_i \nabla_{w_i} \mathcal{L}_i(w_i),~~~ 1\leq j\leq N. \vspace{-0.5em}
    \label{eq:grad_theta}
\end{equation}

Once the gradients are computed\footnote{The method of computing $\nabla_{w_i} \mathcal{L}_i(w_i)$ is exactly equivalent to computing the gradients in case of a single INR network training on image $\widehat{Y}_i$.}, the weights are updated using the conventional gradient descent method or its variants. We next carry out the convergence analysis of our proposed INR training method, leading to a set of convergence upperbounds that analytically verify the soundness of our method and provide practical design insights.
\section{Convergence analysis}
\label{sec:convergence}
Our next goal is to demonstrate the convergence properties of our INR training method. We first note that the general loss function of our interest in~\eqref{eq:loss} can be written as a function of the trainable weights as follows: \vspace{-0.5em}
\begin{equation}
    \label{eq:loss_trainable}
    \mathcal{L}(\Theta) = \sum_{i=1}^{M} \gamma_i \mathcal{L}_i(\sum_{j=1}^{N} \alpha_{ij} \theta_j). \vspace{-0.5em}
\end{equation}
We next make two standard assumptions,  which are common in the machine learning literature~\cite{Hybrid,TranFL,AdaptFL,DinhFL}. 

\begin{assumption}[Smoothness of the Loss Functions]\label{Assup:lossFun1}
    Loss functions $\mathcal{L}_i(.)$  are  $\beta_i$-smooth,~$\forall i$. Mathematically,
     \vspace{-2.mm}
    \begin{equation}
     \exists \beta_i\geq 0: \Vert \mathcal{L}_i(\theta)- \mathcal{L}_i(\theta') \Vert \hspace{-.4mm}\leq \hspace{-.4mm} \beta_i \Vert \theta-\theta' \Vert,~ \hspace{-.8mm}\forall  \theta,\theta'.
    \end{equation}
\end{assumption}   

\begin{assumption}[PL-Condition of the Loss Functions]\label{Assup:lossFun2}
    Loss functions $\mathcal{L}_i(.)$  satisfy the Polyak-Lojasiewicz (PL) condition with constant $\mu_i$,~$\forall i$. Mathematically,\footnote{PL condition is a relaxer condition compared to strong convexity: every strongly convex function satisfies the PL condition but not the other way around. This makes PL condition a practical assumption for analysing broader category of machine learning models that may not possess strongly convex loss functions.}
     \vspace{-2.mm}
    \begin{equation}
     \exists \mu_i \in [0,\beta_i): \Vert \nabla \mathcal{L}_i(\theta) \Vert^2 \hspace{-.4mm}\geq 2\mu_i\left(\mathcal{L}_i(\theta)-\mathcal{L}_i(\theta_i^\star) \right),~ \hspace{-.8mm}\forall  \theta,
    \end{equation}
    where $\theta_i^\star$ is the minimizer of the loss function $\mathcal{L}_i$.
\end{assumption}   
To make the convergence analysis tractable, the convergence is proven with $N=2$ and $M=3$. The extension of the analysis to any $N$ and $M$ is straightforward and omitted for brevity. In this case, we have $w_1 = \theta_1$, $w_2 = \theta_2$, and $w_3 = \alpha_{31} \theta_1 + \alpha_{32} \theta_2$. Also, the loss function is given by
\begin{equation}
    \label{eq:loss_2_3}
    \mathcal{L}(\Theta) = \gamma_1 \mathcal{L}_1(\theta_1) + \gamma_2 \mathcal{L}_2(\theta_2) + \gamma_3 \mathcal{L}_3(\alpha_{31} \theta_1 + \alpha_{32} \theta_2).
\end{equation}
Under the above assumptions, we characterize the convergence behavior of our INR training method in the theorem below.

\begin{theorem}[Convergence Behavior of Constrained INR Training with Linear Combination Method]\label{th:main} Assume that the model training on each INR network  $i\in \{1,2\}$ is conducted using the gradient decent method with step-size of $\eta_i$, satisfying $\eta_i=(\gamma_i\beta_i)^{-1}$. 
Also, let $\theta_i^{(t)}$ denote the  $i$-th INR network $i\in \{1,2\}$ model weight obtained at iteration $t$ of the gradient decent method, for which a uniform upperbound of gradient 
$\|\nabla \mathcal{L}_1(\theta_1^{(t)})\|^2 \leq G_1^2$, $\|\nabla \mathcal{L}_2(\theta_2^{(t)})\|^2 \leq G_2^2$, and $\|\nabla \mathcal{L}_3(w_3^{(t)})\|^2 \leq G_3^2$ hold $\forall t$, with positive constants $G_1^2, G_2^2, G_3^2$. Then, the convergence of the two INR networks for the primary input images and the convergence of the linear combination of the two INR networks for the secondary/third image can be characterized as follows.

\textbf{\textbullet \hspace{1mm} Convergence of the Two INR Networks for the Primary Input Images:}
 The convergence of the loss function $\mathcal{L}_1(.) $ in terms of its optimality gap $\mathcal{L}_1(\theta_1^{(t)}) - \mathcal{L}_1(\theta_1^*)$ at each gradient decent iteration $t$ is given by: \vspace{-0.5em}
\begin{align}
    \label{eq:statement_1}
    \mathcal{L}_1(\theta_1^{(t)}) - \mathcal{L}_1(\theta_1^*) &\leq \left( 1 - \frac{\mu_1}{\beta_1} \right)^t \left(\mathcal{L}_1(\theta_1^{(0)}) - \mathcal{L}_1(\theta_1^*)\right) \nonumber \\
    &\qquad + \frac{\delta_{1,3}}{2 \beta_1\gamma^2_1} \sum_{k=0}^{t-1} \left( 1 - \frac{\mu_1}{\beta_1} \right)^{k},
\end{align}
where $ \delta_{1,3}\triangleq \gamma_3^2 \alpha_{31}^2 G_3^2$, 
which implies the following asymptotic optimality gap
$
   \lim_{t\rightarrow \infty} \mathcal{L}_1(\theta_1^{(t)}) - \mathcal{L}_1(\theta_1^*)= \frac{\delta_{1,3}}{2 \mu_1\gamma^2_1} 
$.

Similarly, for $ \mathcal{L}_2$, we have: \vspace{-0.5em}
\begin{align}
    \label{eq:statement_2}
    \mathcal{L}_2(\theta_2^{(t)}) - \mathcal{L}_1(\theta_2^*) &\leq \left( 1 - \frac{\mu_2}{\beta_2} \right)^t \left(\mathcal{L}_2(\theta_2^{(0)}) - \mathcal{L}_2(\theta_2^*)\right) \nonumber \\
    &\qquad + \frac{\delta_{2,3}}{2 \beta_2\gamma^2_2} \sum_{k=0}^{t-1} \left( 1 - \frac{\mu_2}{\beta_2} \right)^{k},
\end{align}
where $ \delta_{2,3}\triangleq \gamma_3^2 \alpha_{32}^2 G_3^2$, 
which implies the following asymptotic optimality gap
$
   \lim_{t\rightarrow \infty} \mathcal{L}_1(\theta_1^{(t)}) - \mathcal{L}_1(\theta_1^*) = \frac{\delta_{2,3}}{2 \mu_2\gamma^2_2} 
$.

\textbf{\textbullet \hspace{1mm}Convergence of the Linear Combination of the Two INR Networks for the Secondary/Third Image:}
The convergence of $\mathcal{L}_3(.)$ is characterized by the following upperbound \vspace{-0.5em}
\begin{align}\label{eq:last_3}
   \hspace{-2mm} \mathcal{L}_3(w_3^{(t)}) - \mathcal{L}_3(w_3^*) &\leq \left( 1 - \frac{\mu_3}{\beta_3} \right)^t \left(\mathcal{L}_3(w_3^{(0)}) - \mathcal{L}_3(w_3^*)\right) \nonumber \\
    & + \frac{\delta_{1,2,3}}{2 \beta_3} \sum_{k=0}^{t-1} \left( 1 - \frac{\mu_3}{\beta_3} \right)^{k}, \hspace{-3mm} 
\end{align}
which implies the asymptotic optimality gap of $
   \lim_{t\rightarrow \infty}\mathcal{L}_3(w_3^{(t)}) - \mathcal{L}_3(w_3^*) = \frac{\delta_{1,2,3}}{2 \mu_3} .
$ In \eqref{eq:last_3}, $\delta_{1,2,3}$ is given by
\begin{equation} \label{eq:delta_1_2_3}
\begin{aligned}
    &\delta_{1,2,3} \triangleq [1-\gamma_3(\alpha^2_{31}\frac{\beta_3}{\beta_1\gamma_1} +\alpha^2_{32} \frac{\beta_3}{\beta_2\gamma_2})]^2 G_3^2
    \\
    &+\alpha^2_{31} (\frac{\beta_3}{\beta_1})^2 G_1^2+\alpha^2_{32} (\frac{\beta_3}{\beta_2})^2 G_2^2
    \\
    &-2 \alpha_{31} \frac{\beta_3}{\beta_1}[1-\gamma_3(\alpha^2_{31}\frac{\beta_3}{\beta_1\gamma_1} +\alpha^2_{32} \frac{\beta_3}{\beta_2\gamma_2})] \varrho_{1,3}\\
    &-2\alpha_{32} \frac{\beta_3}{\beta_2}[1-\gamma_3(\alpha^2_{31}\frac{\beta_3}{\beta_1\gamma_1} +\alpha^2_{32} \frac{\beta_3}{\beta_2\gamma_2})  ]  \varrho_{2,3}
\\
    &+2\alpha_{31} \frac{\beta^2_3}{\beta_1\beta_2}\alpha_{32} \varrho_{1,2},
\end{aligned}
\end{equation}
where $\varrho_{1,3}, \varrho_{2,3}, \varrho_{1,2}$ are metrics of similarities of the generated gradients during the training period, holding in the following upper and lower bounds (i) $ \langle \nabla_{w_3} \mathcal{L}_3(w_3^{(t)}), \nabla_{\theta_1} \mathcal{L}_1(\theta^{(t)}_1)\rangle \geq  \varrho_{1,3}$, (ii) $ \langle \nabla_{w_3} \mathcal{L}_3(w_3^{(t)}), \nabla_{\theta_2} \mathcal{L}_2(\theta^{(t)}_2)\rangle \geq  \varrho_{2,3}$, (iii) $ \langle \nabla_{\theta_1} \mathcal{L}_1(\theta^{(t)}_1), \nabla_{\theta_2} \mathcal{L}_2(\theta^{(t)}_2)\rangle \leq  \varrho_{1,2}$.

\end{theorem}
\begin{proof}
We first prove the convergence of the two networks for their primary input images and then prove the convergence of the linear combination of the weights of the two networks for the third/secondary image.

\textbf{Convergence of the Two Networks for the Primary Input Images:} In the following, we carry out the proof of convergence for $\mathcal{L}_1(.)$. The convergence of $\mathcal{L}_2(.)$ can be obtained using a similar technique with substituting the indices during the proof steps, and thus omitted for brevity.

The gradient decent update rule used to update the first INR network parameter $\theta_1$  can be written as 
\begin{equation}
    \label{eq:update}
    \begin{gathered}
        \theta_1^{(t+1)} = \theta_1^{(t)} - \eta_{_1} \nabla_{\theta_1} \mathcal{L}(\Theta^{(t)}).
    \end{gathered}
\end{equation}
Also, it is straightforward to verify (see~\cite{Hybrid}) that, using the $\beta_1$-smoothness of $\mathcal{L}_1(.)$ in Assumption~\ref{Assup:lossFun1}, we can get
\begin{equation*}
    \mathcal{L}_1(\theta') \leq \mathcal{L}_1(\theta) + (\nabla \mathcal{L}_1(\theta))^\top (\theta' - \theta) + \frac{\beta_1}{2} \|\theta' - \theta\|^2,~\forall \theta,\theta' .
\end{equation*}
Replacing $\theta'$ with $\theta_1^{(t+1)}$ and $\theta$ with $\theta^{(t)}$ in the above expression implies the following inequality \vspace{-0.5em}
\begin{align}
    \mathcal{L}_1(\theta_1^{(t+1)}) &\leq \mathcal{L}_1(\theta_1^{(t)}) + \left( \nabla \mathcal{L}_1(\theta_1^{(t)}) \right)^\top (\theta_1^{(t+1)} - \theta_1^{(t)}) \nonumber \\
    &\qquad+ \frac{\beta_1}{2} \|\theta_1^{(t+1)} - \theta_1^{(t)}\|^2. \vspace{-0.5em}
\end{align}
Substituting the update rule of \eqref{eq:update} in the above expression and rearranging the terms yields \vspace{-0.5em}
\begin{align}
    \label{eq:smooth_1}
    \mathcal{L}_1(\theta_1^{(t+1)}) &\leq \mathcal{L}_1(\theta_1^{(t)}) - \eta_{_1} \left( \nabla \mathcal{L}_1(\theta_1^{(t)}) \right)^\top \nabla_{\theta_1} \mathcal{L}(\Theta^{(t)}) \nonumber \\
    &\qquad+ \frac{\beta_1 \eta_{_1}^2}{2} \|\nabla_{\theta_1} \mathcal{L}(\Theta^{(t)})\|^2.
\end{align}
We next use the fact that the gradient of the loss in \eqref{eq:loss_2_3} with respect to $\theta_1$, i.e., term $\nabla_{\theta_1} \mathcal{L}(\Theta^{(t)})$ in \eqref{eq:smooth_1}, can be split into the scaled gradient of the local loss function with respect to the local model weight $\nabla_{\theta_1} \mathcal{L}_1(\theta_1)$ and an error term. In particular, using the chain rule of differentiation, we can get
\begin{align} \label{eq:upErr_1}
    \nabla_{\theta_1} \mathcal{L}(\Theta) &= \gamma_1 \nabla_{\theta_1} \mathcal{L}_1(\theta_1) + \gamma_3 \alpha_{31} \nabla_{w_3} \mathcal{L}_3(w_3),
\end{align}
which implies \vspace{-0.5em}
\begin{align}\label{eq:trueANDerror}
    \nabla_{\theta_1} \mathcal{L}(\Theta) &= \gamma_1 \nabla_{\theta_1} \mathcal{L}_1(\theta_1) + e_1,
\end{align}
where the error is given by
    $e_1 = \gamma_3 \alpha_{31} \nabla_{w_3} \mathcal{L}_3(w_3).$

Substituting \eqref{eq:trueANDerror} in \eqref{eq:smooth_1}, while defining $e^{(t)}_1 = \gamma_3 \alpha_{31} \nabla_{w_3} \mathcal{L}_3(w^{(t)}_3)$, where $w^{(t)}_3=\alpha_{31} \theta^{(t)}_1 + \alpha_{32} \theta^{(t)}_2$, gives us the following inequality
\begin{align}
    \label{eq:smooth_2}
    \mathcal{L}_1(\theta_1^{(t+1)}) &\leq \mathcal{L}_1(\theta_1^{(t)})  - \eta_{_1} \Big(\gamma_1 \Vert\nabla_{\theta_1} \mathcal{L}_1(\theta_1) \Vert^2 \nonumber \\
    &+ \big( \nabla \mathcal{L}_1(\theta_1^{(t)}) \big)^\top e_1^{(t)}\Big) + \frac{\beta_1 \eta_{_1}^2}{2}\Big( \Vert\gamma_1 \nabla_{\theta_1} \mathcal{L}_1(\theta_1)\Vert^2 
    \nonumber \\
    &+ 2\big(\gamma_1 \nabla \mathcal{L}_1(\theta_1^{(t)}) \big)^\top e_1^{(t)} + \Vert e^{(t)}_1\Vert^2  \Big).
\end{align}    
With some rearranging of the terms, we can get
\begin{align}
    \mathcal{L}_1(\theta_1^{(t+1)}) &\leq \mathcal{L}_1(\theta_1^{(t)}) + \| \nabla \mathcal{L}_1(\theta_1^{(t)}) \|^2 \left( \frac{\beta_1 \eta_{_1}^2\gamma_1^2}{2} - \eta_{_1}\gamma_1 \right) \nonumber \\
    &+ \left( \nabla \mathcal{L}_1(\theta_1^{(t)}) \right)^\top e_1^{(t)} \left( \beta_1 \eta_{_1}^2\gamma_1 - \eta_{_1} \right) \nonumber \\
    &+ \frac{\beta_1 \eta_{_1}^2}{2} \| e_1^{(t)} \|^2.
\end{align}   
Choosing the learning rate $\eta_{_1} = (\beta_1 \gamma_1)^{-1}$, the above inequality simplifies to 
\begin{equation}
    \label{eq:smooth_3}
  \hspace{-4mm}  \resizebox{.92\linewidth}{!}{$
    \mathcal{L}_1(\theta_1^{(t+1)}) \leq \mathcal{L}_1(\theta_1^{(t)})  - \frac{1}{2 \beta_1} \| \nabla \mathcal{L}_1(\theta_1^{(t)}) \|^2 + \frac{1}{2 \beta_1\gamma_1^2} \| e_1^{(t)} \|^2.$} \hspace{-4mm}
\end{equation}
Using the PL-condition in Assumption~\ref{Assup:lossFun2} in the above inequality implies
\begin{equation}
    \label{eq:smooth_3}
   \hspace{-3mm}\resizebox{.92\linewidth}{!}{$ \mathcal{L}_1(\theta_1^{(t+1)}) \leq \mathcal{L}_1(\theta_1^{(t)})  - \frac{\mu_1}{\beta_1} \left(\mathcal{L}_1(\theta_1^{(t)}) - \mathcal{L}_1(\theta_1^*)\right) + \frac{1}{2 \beta_1\gamma^2_1} \| e_1^{(t)} \|^2.$}\hspace{-3mm}
\end{equation}
Subtracting $\mathcal{L}_1(\theta_1^*)$ from both hand sides of the above inequality yields the following one step rule on the optimality gap
\begin{align}
    \label{eq:smooth_3}
    &\mathcal{L}_1(\theta_1^{(t+1)}) - \mathcal{L}_1(\theta_1^*) \leq \left( 1 - \frac{\mu_1}{\beta_1} \right) \left(\mathcal{L}_1(\theta_1^{(t)}) - \mathcal{L}_1(\theta_1^*)\right) \nonumber \\
    &\qquad\qquad\qquad\qquad\qquad + \frac{1}{2 \beta_1 \gamma^2_1} \| e_1^{(t)} \|^2.
\end{align}
Recursive expansion of the above update rule, assuming the initial weights $\theta_1^{(0)}$, gives us
\begin{align}
    \label{eq:smooth_one_step}
   \hspace{-2mm} \mathcal{L}_1(\theta_1^{(t)}) - \mathcal{L}_1(\theta_1^*) &\leq \left( 1 - \frac{\mu_1}{\beta_1} \right)^t \left(\mathcal{L}_1(\theta_1^{(0)}) - \mathcal{L}_1(\theta_1^*)\right) \nonumber \\
    & + \frac{1}{2 \beta_1\gamma^2_1} \sum_{k=0}^{t-1} \left( 1 - \frac{\mu_1}{\beta_1} \right)^{k} \| e_1^{(t-k)} \|^2. \hspace{-3mm} \vspace{-1em} 
\end{align}
Note that 
\begin{align}
    \label{eq:error_1_norm}
    \| e_1^{(t)} \|^2 &= \gamma_3^2 \alpha_{31}^2 \|\nabla \mathcal{L}_3(w_3^{(t)})\|^2\leq \delta_{1,3}\triangleq \gamma_3^2 \alpha_{31}^2 G_3^2.
\end{align}
    Thus, the convergence bound in \eqref{eq:smooth_one_step} can be written as follows:
\begin{align}
    \label{eq:smooth_6}
    \mathcal{L}_1(\theta_1^{(t)}) - \mathcal{L}_1(\theta_1^*) &\leq \left( 1 - \frac{\mu_1}{\beta_1} \right)^t \left(\mathcal{L}_1(\theta_1^{(0)}) - \mathcal{L}_1(\theta_1^*)\right) \nonumber \\
    &\qquad + \frac{\delta_{1,3}}{2 \beta_1\gamma^2_1} \sum_{k=0}^{t-1} \left( 1 - \frac{\mu_1}{\beta_1} \right)^{k}.
\end{align}
Noting that $\mu_1<\beta_1$, which implies $\lim_{t\rightarrow \infty}\sum_{k=0}^{t-1} \left( 1 - \frac{\mu_1}{\beta_1} \right)^{k} =\frac{\beta_1}{\mu_1}$, taking the $\lim_{t\rightarrow \infty}$ from both hand sides of the above bound yields 
$
   \lim_{t\rightarrow \infty} \mathcal{L}_1(\theta_1^{(t)}) - \mathcal{L}_1(\theta_1^*) = \frac{\delta_{1,3}}{2 \mu_1\gamma^2_1} .
$

\textbf{Convergence of the Linear Combination of the Two INR Networks for the Secondary/Third Image:} We next conduct the analysis for $\mathcal{L}_3(.)$. Note that $\mathcal{L}_3(.)$ is not trained directly; instead it is evaluated on $w_3=\alpha_{31} \theta_1 + \alpha_{32} \theta_2$, which is the linear combination of the two INR network weights. Thus, at gradient descent iteration $t$, we have $w^{(t)}_3=\alpha_{31} \theta^{(t)}_1 + \alpha_{32} \theta^{(t)}_2$. Thus, the difference between two consecutive values of $w_3$ during the gradient descent iterations can be written as follows:
\begin{equation}
    \label{eq:update_w}
    \hspace{-2mm}
    \resizebox{.92\linewidth}{!}{$ 
    \begin{gathered}
    w_3^{(t+1)} - w_3^{(t)} = \alpha_{31} \left( \theta_1^{(t+1)} - \theta_1^{(t)} \right) + \alpha_{32} \left( \theta_2^{(t+1)} - \theta_2^{(t)} \right) \\
   \Rightarrow w_3^{(t+1)} = w_3^{(t)} -  \left( \alpha_{31} \eta_{_1}\nabla_{\theta_1} \mathcal{L}(\Theta^{(t)}) + \alpha_{32} \eta_{_2}\nabla_{\theta_2} \mathcal{L}(\Theta^{(t)}) \right).
   \end{gathered}
   $}\hspace{-3.5mm}
\end{equation}
Consequently, similar to \eqref{eq:upErr_1} and~\eqref{eq:trueANDerror}, the update rule for $w_3$ can be written as the following (hypothetical) gradient decent update with any choice of $\eta_3$:
\begin{equation}
    \label{eq:grad_w_3}
    w_3^{(t+1)} = w_3^{(t)} - \eta_{_3} \left( \nabla_{w_3} \mathcal{L}_3(w_3^{(t)}) + e_3^{(t)} \right),
\end{equation}
where, the error term $e_3^{(t)}$ can be computed as follows:
\begin{equation}\label{eq:e_3}
\hspace{-4mm}
 \resizebox{.92\linewidth}{!}{$ 
\begin{aligned}
    & e_3^{(t)} =  -\nabla_{w_3} \mathcal{L}_3(w_3^{(t)})+\alpha_{31} \frac{\eta_{_1}}{\eta_{_3}}\nabla_{\theta_1} \mathcal{L}(\Theta^{(t)}) + \alpha_{32} \frac{\eta_{_2}}{\eta_{_3}}\nabla_{\theta_2} \mathcal{L}(\Theta^{(t)})  \\
    & \Rightarrow e_3^{(t)} =   -\nabla_{w_3} \mathcal{L}_3(w_3^{(t)})+\alpha_{31} \frac{\eta_{_1}}{\eta_{_3}}\gamma_1 \nabla_{\theta_1} \mathcal{L}_1(\theta^{(t)}_1) +\alpha_{31} \frac{\eta_{_1}}{\eta_{_3}} e^{(t)}_1 +\\
    &~~\alpha_{32} \frac{\eta_{_2}}{\eta_{_3}}\gamma_2 \nabla_{\theta_2} \mathcal{L}_2(\theta^{(t)}_2)+\alpha_{32} \frac{\eta_{_2}}{\eta_{_3}} e^{(t)}_2
     \\
     &\Rightarrow e_3^{(t)} = -\nabla_{w_3} \mathcal{L}_3(w_3^{(t)})+\alpha_{31} \frac{\eta_{_1}}{\eta_{_3}}\gamma_1 \nabla_{\theta_1} \mathcal{L}_1(\theta^{(t)}_1)  \\
    &~~+ \alpha_{32} \frac{\eta_{_2}}{\eta_{_3}}\gamma_2 \nabla_{\theta_2} \mathcal{L}_2(\theta^{(t)}_2) ~~+\alpha_{31}\frac{\eta_{_1}}{\eta_{_3}} \gamma_3 \alpha_{31} \nabla_{w_3} \mathcal{L}_3(w^{(t)}_3)  \\
    &~~+\alpha_{32} \frac{\eta_{_2}}{\eta_{_3}}\gamma_3 \alpha_{32} \nabla_{w_3} \mathcal{L}_3(w^{(t)}_3) \\
     &\Rightarrow e_3^{(t)} = -[1-\gamma_3(\alpha^2_{31}\frac{\eta_{_1}}{\eta_{_3}} +\alpha^2_{32} \frac{\eta_{_2}}{\eta_{_3}}) 
    ]\nabla_{w_3} \mathcal{L}_3(w_3^{(t)}) \\
    &~~+\alpha_{31} \frac{\eta_{_1}}{\eta_{_3}}\gamma_1 \nabla_{\theta_1} \mathcal{L}_1(\theta^{(t)}_1) 
    + \alpha_{32} \frac{\eta_{_2}}{\eta_{_3}}\gamma_2 \nabla_{\theta_2} \mathcal{L}_2(\theta^{(t)}_2).
    \end{aligned}
     $} \hspace{-4mm}
\end{equation}
With choosing $\eta_3=(\beta_3)^{-1}$ and following the same steps in the convergence proof of $\mathcal{L}_1(.)$, we can obtain the following optimality gap for $\mathcal{L}_3(.)$
\begin{align}
    \label{eq:smooth_L_3}
   \hspace{-2mm} \mathcal{L}_3(w_3^{(t)}) - \mathcal{L}_3(w_3^*) &\leq \left( 1 - \frac{\mu_3}{\beta_3} \right)^t \left(\mathcal{L}_3(w_3^{(0)}) - \mathcal{L}_3(w_3^*)\right) \nonumber \\
    & + \frac{1}{2 \beta_3} \sum_{k=0}^{t-1} \left( 1 - \frac{\mu_3}{\beta_3} \right)^{k} \| e_3^{(t-k)} \|^2. \hspace{-3mm} 
\end{align}
Note that since $\eta_1=(\beta_1 \gamma_1)^{-1}, \eta_2=(\beta_2 \gamma_2)^{-1}, \eta_3=(\beta_3)^{-1}$, assuming $1-\gamma_3(\alpha^2_{31}\frac{\beta_3}{\beta_1\gamma_1} +\alpha^2_{32} \frac{\beta_3}{\beta_2\gamma_2})\geq 0\Rightarrow \gamma_3 \leq \frac{1}{\alpha^2_{31}\frac{\beta_3}{\beta_1\gamma_1} +\alpha^2_{32} \frac{\beta_3}{\beta_2\gamma_2}}$, using the result in~\eqref{eq:e_3},  we can get
\begin{equation}
\hspace{-3mm}
 \resizebox{.93\linewidth}{!}{$ 
\begin{aligned}
    \label{eq:error_1_norm}
    &\| e_3^{(t)} \|^2 = [1-\gamma_3(\alpha^2_{31}\frac{\beta_3}{\beta_1\gamma_1} +\alpha^2_{32} \frac{\beta_3}{\beta_2\gamma_2}) 
    ]^2 \Vert \nabla_{w_3} \mathcal{L}_3(w_3^{(t)})\Vert^2 \\
    &+\alpha^2_{31} (\frac{\beta_3}{\beta_1\gamma_1})^2\gamma^2_1 \Vert\nabla_{\theta_1} \mathcal{L}_1(\theta^{(t)}_1)\Vert^2 + \alpha^2_{32} (\frac{\beta_3}{\beta_2\gamma_2})^2\gamma^2_2 \Vert\nabla_{\theta_2} \mathcal{L}_2(\theta^{(t)}_2)\Vert^2\\
    &
    - 2\alpha_{31} \frac{\beta_3}{\beta_1\gamma_1}\gamma_1[1-\gamma_3(\alpha^2_{31}\frac{\beta_3}{\beta_1\gamma_1} +\alpha^2_{32} \frac{\beta_3}{\beta_2\gamma_2}) 
    ](\nabla_{w_3} \mathcal{L}_3(w_3^{(t)}))^\top  \nabla_{\theta_1} \mathcal{L}_1(\theta^{(t)}_1) 
    \\
    &
    -2\alpha_{32} \frac{\beta_3}{\beta_2\gamma_2}\gamma_2[1-\gamma_3(\alpha^2_{31}\frac{\beta_3}{\beta_1\gamma_1} +\alpha^2_{32} \frac{\beta_3}{\beta_2\gamma_2}) 
    ](\nabla_{w_3} \mathcal{L}_3(w_3^{(t)}))^\top \nabla_{\theta_2} \mathcal{L}_2(\theta^{(t)}_2)
    \\
    &
    +2\alpha_{31} \frac{\beta_3}{\beta_1\gamma_1}\gamma_1\alpha_{32} \frac{\beta_3}{\beta_2\gamma_2}\gamma_2  (\nabla_{\theta_1} \mathcal{L}_1(\theta^{(t)}_1))^\top \nabla_{\theta_2} \mathcal{L}_2(\theta^{(t)}_2)\leq \delta_{1,2,3} ,
\end{aligned}
$}\hspace{-3mm}
\end{equation}
where
\begin{equation}
\hspace{-3mm}
\begin{aligned}
    &\delta_{1,2,3} \triangleq [1-\gamma_3(\alpha^2_{31}\frac{\beta_3}{\beta_1\gamma_1} +\alpha^2_{32} \frac{\beta_3}{\beta_2\gamma_2})]^2 G_3^2
    \\
    &
    +\alpha^2_{31} (\frac{\beta_3}{\beta_1})^2 G_1^2+\alpha^2_{32} (\frac{\beta_3}{\beta_2})^2 G_2^2
    \\
    &-2 \alpha_{31} \frac{\beta_3}{\beta_1}[1-\gamma_3(\alpha^2_{31}\frac{\beta_3}{\beta_1\gamma_1} +\alpha^2_{32} \frac{\beta_3}{\beta_2\gamma_2})] \varrho_{1,3}\\
    &-2\alpha_{32} \frac{\beta_3}{\beta_2}[1-\gamma_3(\alpha^2_{31}\frac{\beta_3}{\beta_1\gamma_1} +\alpha^2_{32} \frac{\beta_3}{\beta_2\gamma_2})  ]  \varrho_{2,3}\\
    &
+2\alpha_{31} \frac{\beta^2_3}{\beta_1\beta_2}\alpha_{32} \varrho_{1,2}.
\end{aligned}
\end{equation}
Using the above result in \eqref{eq:smooth_L_3}, yields
\begin{align}
   \hspace{-2mm} \mathcal{L}_3(w_3^{(t)}) - \mathcal{L}_3(w_3^*) &\leq \left( 1 - \frac{\mu_3}{\beta_3} \right)^t \left(\mathcal{L}_3(w_3^{(0)}) - \mathcal{L}_3(w_3^*)\right) \nonumber \\
    & + \frac{\delta_{1,2,3}}{2 \beta_3} \sum_{k=0}^{t-1} \left( 1 - \frac{\mu_3}{\beta_3} \right)^{k}, \hspace{-3mm} 
\end{align}
which implies the asymptotic optimality gap of $
   \lim_{t\rightarrow \infty}\mathcal{L}_3(w_3^{(t)}) - \mathcal{L}_3(w_3^*) = \frac{\delta_{1,2,3}}{2 \mu_3} .
$
\end{proof}


\begin{remark}[Interpretation of the Results and the Upperbounds on the Primary Images]
Inspecting the convergence bounds of the original INR networks trained on the primary images given by~\eqref{eq:statement_1} and~\eqref{eq:statement_2}, we can observe the exponential (also called linear) convergence rate of the loss function, which is the best convergence rate that can be achieved via gradient decent method. Furthermore, the asymptotic optimality gaps on $\mathcal{L}_1$ and $\mathcal{L}_2$, which are functions of $\delta_{1,3}$ and $\delta_{2,3}$, imply that introducing the loss function of the third image during the model training of each of the primary images induces a finite optimality gap for the original images. Inspecting the expressions for $\delta_{1,3}$ $\delta_{2,3}$, it can be construed that this optimality gap is a function of the weight/importance of the third image during model training (i.e., constant $\gamma_3$ and combiner weights $\alpha_{3,1}$ and $\alpha_{3,2}$).
\end{remark}

\begin{remark}[Interpretation of the Results and the Upperbounds on the Secondary Image]
Inspecting the convergence bound of the linear combination of the INR networks on the secondary image given by~\eqref{eq:last_3}, we still observe the exponential convergence rate of the loss function. Nevertheless, the structure of the asymptotic optimally gap on $\mathcal{L}_3$, which is a function of $\delta_{1,2,3}$ is different from those on $\mathcal{L}_1$ and $\mathcal{L}_2$. Inspecting the expressions for $\delta_{1,2,3}$, the bound has an intuitive implication: since the third image uses the gradients generated over the other two images, the convergence of the loss function $\mathcal{L}_3$ is a function of the similarity of those generated with that of the hypothetical gradient that describes the actual gradient of the image over the combined weights, i.e., $\nabla_{w_3} \mathcal{L}_3(w_3^{(t)})$. The bound captures this similarity through the inner product between these gradients encoded in $\varrho_{1,3}$ and $\varrho_{2,3}$. Note that as these similarities grow (i.e., the $3^{rd}$ image has more similarity to the other two images), the values of $\varrho_{1,3}$ and $\varrho_{2,3}$ will grow, which will result in a smaller optimality gap (note the terms with negative signs in the expression of $\delta_{1,2,3}$ in~\eqref{eq:delta_1_2_3}), and thus a better convergence behavior for the $3^{rd}$ image.
\end{remark}

\section{Benchmarking}
\label{sec:benchmarking}

The proposed training method is validated on a few image datasets, \textit{viz.}, Kodak image suite \cite{Kodak}, ImageNet \cite{ImageNet}, and CIFAR-10 \cite{CIFAR10}. The height ($H$), width ($W$), and channels ($C$) of the images in each dataset are necessary to compute the compression ratios. Table \ref{tab:dataset_info} summarizes the image sizes of the three datasets used. The model topology used is a fully connected network with sinusoidal activation functions (SIREN) \cite{Sitzmann2020}. All the networks in this section have the same number of neurons in each hidden layer. The number of bits required for storing each weight is represented by $B_P$. In this work, the weights are stored as $16$-bit floating point numbers. All the PSNR results are obtained with $B_P=16$ as the degradation in PSNR is $< 0.2\%$ when quantizing from $32$-bit to $16$-bit floating point numbers.\vspace{-0.5em}

\begin{table}[h]
    \renewcommand{\arraystretch}{1.2}
    \centering
    \vspace{-0.75em}
    \caption{Image sizes of the datasets being used.}
    \vspace{-0.5em}
    \label{tab:dataset_info}
    \begin{tabular}{|c|c|c|c|}
        \hline
        \hline
 \rowcolor{Gray}        \textbf{Dataset} & $\mathbf{H}$ & $\mathbf{W}$ & $\mathbf{C}$ \\
        \hline
        \hline
        \textbf{Kodak} & $512$ & $768$ & $3$ \\
        \hline
        \textbf{ImageNet} & $256$ & $256$ & $3$ \\
        \hline
        \textbf{CIFAR-10} & $32$ & $32$ & $3$ \\
        \hline
        \hline
    \end{tabular}
\end{table}

\vspace{-2em}
\subsection{Kodak Image Suite}
\begin{figure}[htbp]
\vspace{-1mm}
    \centering
    \includegraphics[scale=0.8]{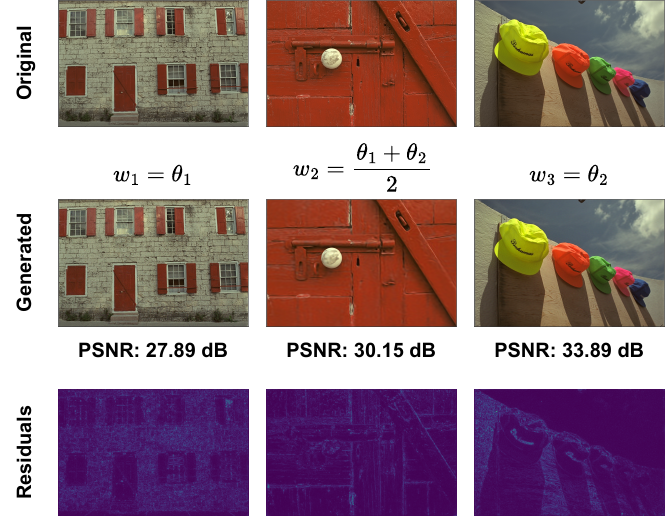} \vspace{-2em}
    \caption{Generated images using the proposed training method on the first three images of the Kodak dataset along with their peak signal-to-noise ratio (PSNR) and residuals.} \vspace{-1em}
    \label{fig:kodak_first_3}
    \vspace{2mm}
\end{figure}

As an initial proof of concept, the proposed training method is tested on the first three images of the Kodak dataset. Figure \ref{fig:kodak_first_3} shows the generated images after training along with their residuals. The network used has $4$ hidden layers with $256$ neurons each and is trained for $5000$ epochs. Two sets of weights, $\theta_1$ and $\theta_2$, are used to generate the three images ($N=2$, $M=3$). The linear combination used is $w_1 = \theta_1$, $w_2 = (\theta_1 + \theta_2)/2$, and $w_3 = \theta_2$. As shown in Fig. \ref{fig:kodak_first_3}, the proposed training method is able to generate the three images with high fidelity. The peak signal-to-noise ratio (PSNR) of the generated images is $30.64$ dB on average.
\begin{figure*}[htbp]
    \centering
    \includegraphics[scale=0.9]{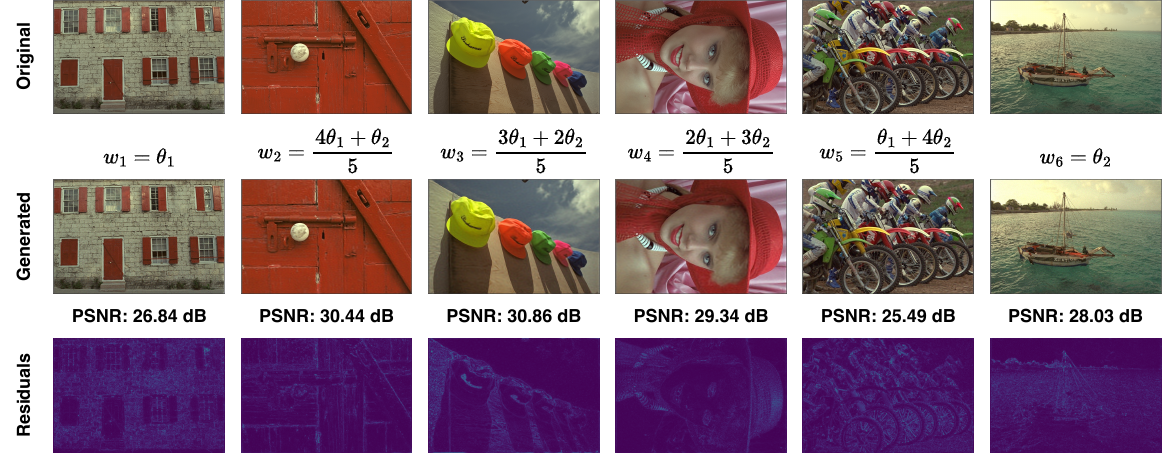} \vspace{-0.75em}
    \caption{Generated images using the proposed training method on the first six images of the Kodak dataset and their corresponding residuals. The BPP in this case is $0.9$.}
    \label{fig:kodak_first_6}
    \vspace{-1mm}
\end{figure*}

The idea is extended for $N=2$ and $M=6$. Figure \ref{fig:kodak_first_6} shows the generated images after training for $5000$ epochs with the same network structure as the previous case. The Kodak dataset has images in both portrait and landscape orientations. The images in portrait orientation are rotated to make the image size consistent. The average PSNR of the generated images is $28.50$ dB. The PSNR is lower than in the previous case as more images are generated using the same number of weights, network structure, and iterations. From the residuals, it is evident that the network cannot faithfully generate high-frequency details in the images, which is the case with most INR networks. The models can be made to learn the high-frequency details better by changing the weight initialization distribution or using predetermined positional encodings. Such analysis will be left for our future work.

The number of parameters in the network used is about 200 K, requiring about 777 KB of memory to store two sets of weights. This results in a bit rate of $0.9$ bits per pixel (BPP) for generating $6$ images\footnote{A standard image is represented using 8 bits per pixel. Therefore, a BPP of $0.9$ implies that the image is compressed by a factor of $8.9$.}. The BPP can be reduced by using a smaller network. For example, using a network with $4$ hidden layers and $170$ neurons per layer instead of $256$ gives a bit rate of $0.4$ BPP. However, the same BPP can be achieved with various network depths. Table \ref{tab:network_depth_4} shows the PSNR of the generated images for different network depths. All networks are trained for $5000$ epochs. The PSNR is higher for deeper networks as they can model the images better. However, the PSNR is not significantly different for network depths greater than $7$. The results in Table \ref{tab:network_depth_4} are for generating $6$ images using $2$ sets of weights. The Kodak dataset has $24$ images, which were split into $4$ sets of $6$ images each. The reported PSNR in Table \ref{tab:network_depth_4} is the average PSNR of the $24$ images. The number of parameters, $P$, is given by (\ref{eq:num_params}), and the bit rate, BPP, is given by (\ref{eq:BPP})\footnote{The values of $H$, $W$, and $C$ are given in Table \ref{tab:dataset_info} and $B_P=16$.}.

\begin{table}[t]
    \renewcommand{\arraystretch}{1.2}
    \centering
    \caption{PSNR of the generated images for different network depths achieving a bit rate of $0.4$ BPP for $N=2$ and $M=6$.}
    \label{tab:network_depth_4}
    \begin{tabular}{|c|c|c|c|c|}
    \hline
    \hline
   \rowcolor{Gray}  \Centerstack{\textbf{\# Layers} ($l$)} & \Centerstack{\textbf{Neurons per layer} (n)} & \Centerstack{\textbf{\# Params} (P)} & \textbf{BPP} & \textbf{PSNR (dB)} \\
    \hline
    \hline
    $2$ & $294$ & $88497$ & $0.400$ & $24.96$ \\
    \hline
    $4$ & $170$ & $88233$ & $0.399$ & $27.32$ \\
    \hline
    $6$ & $132$ & $88575$ & $0.400$ & $28.06$ \\
    \hline
    $7$ & $120$ & $87843$ & $0.397$ & $28.23$ \\
    \hline
    $8$ & $112$ & $89267$ & $0.404$ & $28.50$ \\
    \hline
    $10$ & $98$ & $87909$ & $0.397$ & $28.46$ \\
    \hline
    \hline
    \end{tabular}
    \vspace{-1mm}
\end{table}

\begin{equation}
    \label{eq:num_params}
    P = 3 \cdot n + (l-1) \cdot (n^2 + n) + (n + 1) \cdot 3 \vspace{-0.25em}
\end{equation}
\begin{equation}
    \label{eq:BPP}
    \text{BPP} = \frac{N \cdot P \cdot B_P}{M \cdot H \cdot W \cdot C}
\end{equation}

The same experiment is repeated for a different target bit rate of $0.2$ BPP, as shown in Table \ref{tab:network_depth_2}. Similar to the previous case, the PSNR is higher for deeper networks and saturates for network depths greater than $7$. For the rest of this work, all the networks used are $8$ layers deep.

\begin{table}[htbp]
    \vspace{-.7em}
    \renewcommand{\arraystretch}{1.2}
    \centering
    \caption{PSNR of the generated images for different network depths achieving a bit rate of $0.2$ BPP for $N=2$ and $M=6$.}
    \label{tab:network_depth_2}
    \begin{tabular}{|c|c|c|c|c|}
    \hline
    \hline
    \rowcolor{Gray}  \Centerstack{\textbf{\# Layers} ($l$)} & \Centerstack{\textbf{Neurons per layer} (n)} & \Centerstack{\textbf{\# Params} (P)} & \textbf{BPP} & \textbf{PSNR (dB)} \\
    \hline
    \hline
    $2$ & $206$ & $43881$ & $0.198$ & $24.18$ \\
    \hline
    $4$ & $120$ & $44283$ & $0.200$ & $25.76$ \\
    \hline
    $6$ & $92$ & $43335$ & $0.196$ & $26.19$ \\
    \hline
    $7$ & $84$ & $43347$ & $0.196$ & $26.24$ \\
    \hline
    $8$ & $78$ & $43605$ & $0.197$ & $26.42$ \\
    \hline
    $10$ & $70$ & $45153$ & $0.204$ & $26.62$ \\
    \hline
    \hline
    \end{tabular}
    \vspace{-.2em}
\end{table}

\begin{figure}[htbp]
    \centering
    \includegraphics[scale=0.425]{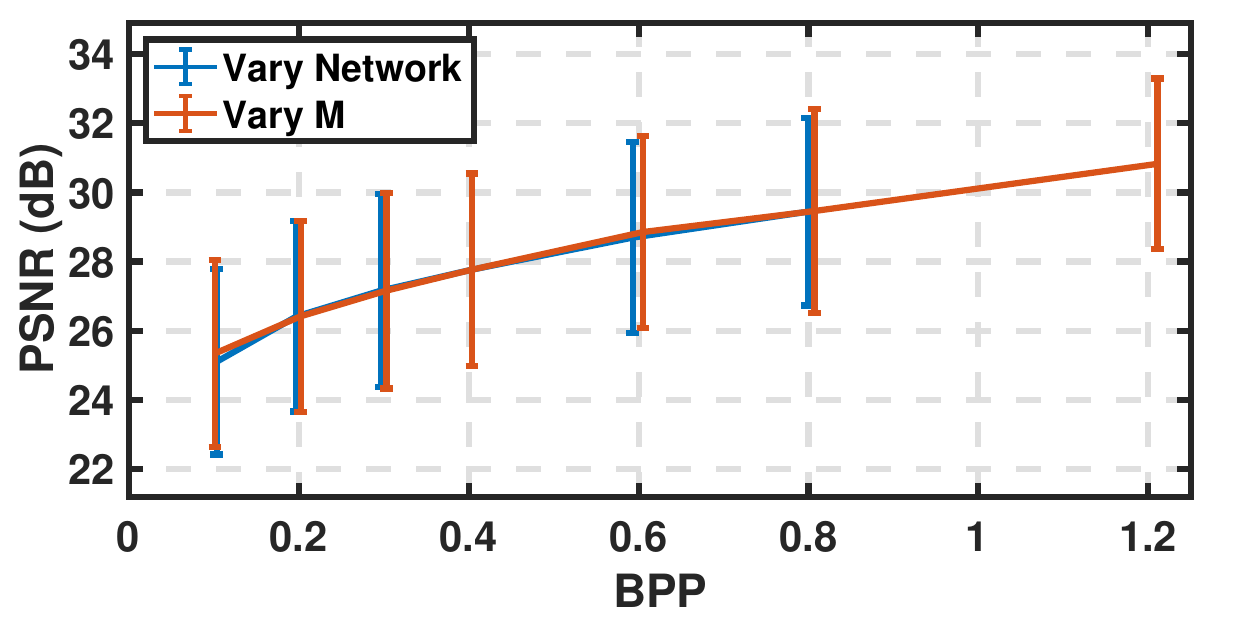}
    \vspace{-1em}
    \caption{Rate-distortion plot for different bit rates achieved by changing the network architecture and the number of images to reconstruct ($M$). The error bars represent the standard deviation of the PSNR for all the images in the dataset.}
    \label{fig:network_image_change} \vspace{-1mm}
\end{figure}

There are two ways of changing the bit rate of the generated images: (i) by changing the network architecture and (ii) by changing the number of images to reconstruct. Figure \ref{fig:network_image_change} shows the rate-distortion curve for various bit rates for the two methods. The networks are all $8$ layers deep and are trained for $5000$ epochs. The rate-distortion curve is obtained by varying the number of neurons per layer from $56$ to $158$ and the number of images to reconstruct from $3$ to $24$. In the case of varying $M$, the non-trainable weights are computed according to (\ref{eq:vary_M_weights}).
\begin{equation}
    \label{eq:vary_M_weights}
    w_i = \frac{M-i}{M-1}\theta_1 + \frac{i-1}{M-1}\theta_2 \vspace{-1em}
\end{equation}

As evident from Fig. \ref{fig:network_image_change}, the PSNR in both cases are very similar to each other. For the rest of this work, the bit rate is changed by changing the number of images to reconstruct. It is to be noted that the proposed method is able to represent all $24$ images in the Kodak dataset with just two sets of weights, resulting in a BPP of $0.1$ and achieving an average PSNR of $25.34$ dB.

\subsection{CIFAR-10}
\begin{figure}[htbp]
    \centering
    \includegraphics[scale=0.425]{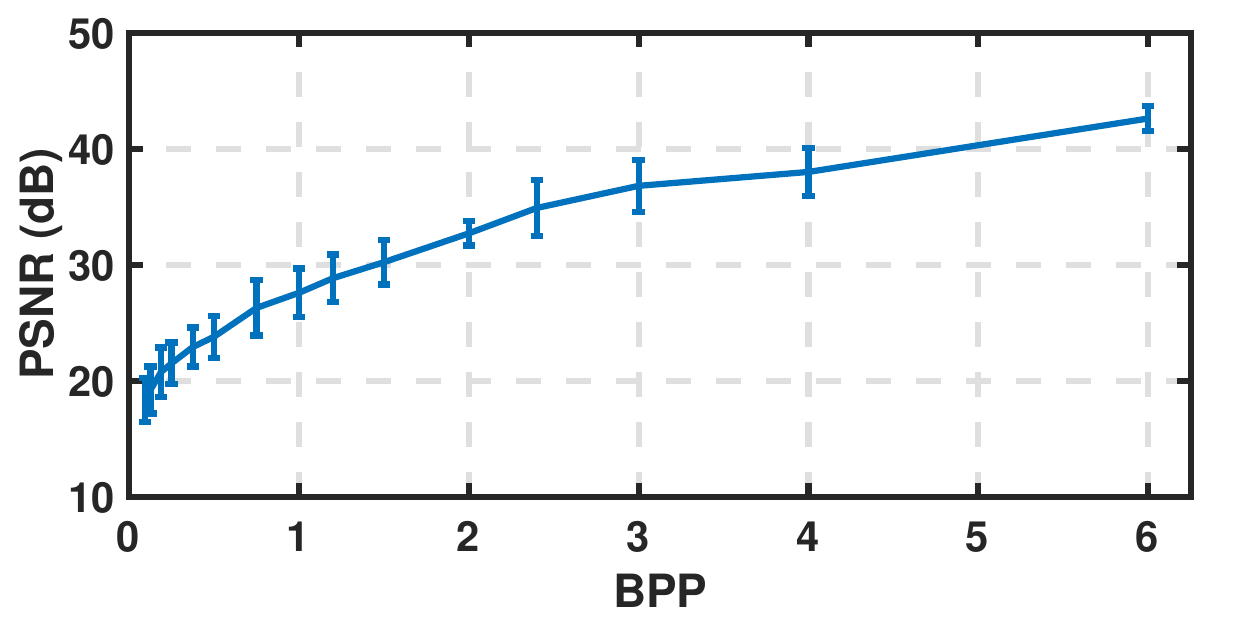}
    \vspace{-1em}
    \caption{Rate-distortion plot for different bit rates achieved by changing the number of images to reconstruct ($M$) for the CIFAR-10 dataset.} \vspace{-1.5em}
    \label{fig:cifar10}
\end{figure}

Images in the test set of CIFAR-10 are compressed using the proposed method. The network used is $4$ layers deep with $18$ neurons per layer. The network is trained for $10000$ epochs. The rate-distortion curve for the CIFAR-10 dataset is shown in Fig. \ref{fig:cifar10}. The lowest bit rate in Fig. \ref{fig:cifar10} is $0.094$ BPP, which is achieved by reconstructing $128$ images using $2$ sets of weights, achieving an average PSNR of $18.35$ dB. 

\subsection{ImageNet}
\begin{figure}[htbp]
    \centering
    \includegraphics[scale=0.425]{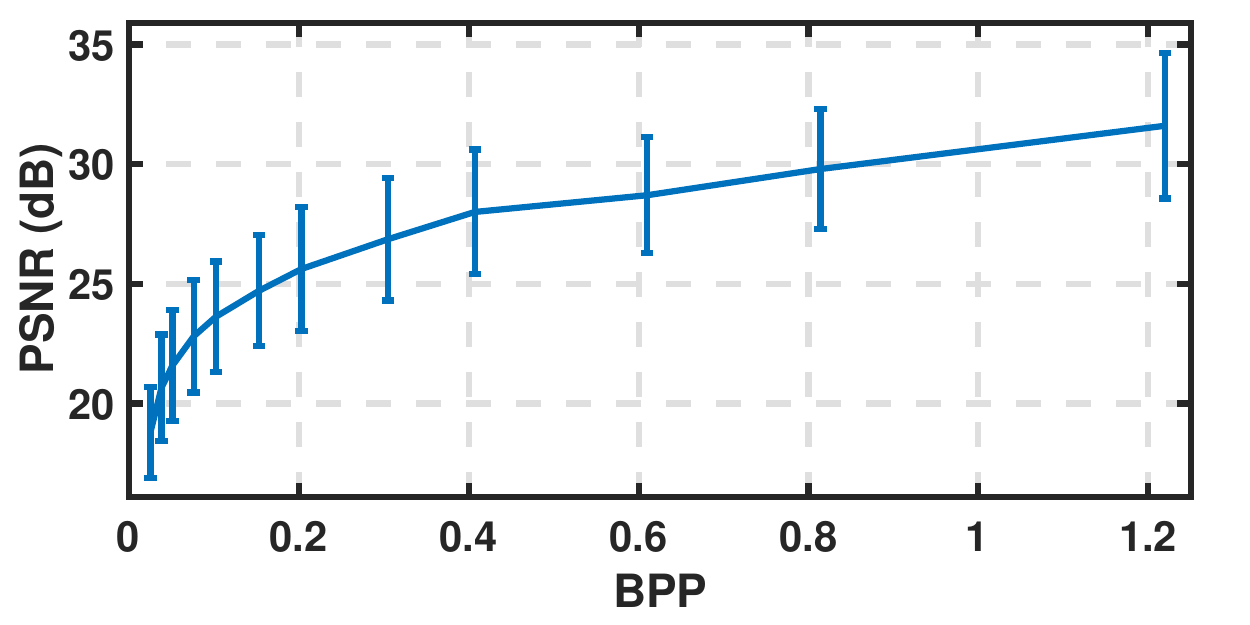}
             \vspace{-1em}
    \caption{Rate-distortion plot for different bit rates achieved by changing the number of images to reconstruct ($M$) for the ImageNet dataset.} \vspace{-1em}
    \label{fig:imagenet}
\end{figure}

The proposed method is also validated by training on the first $20$ classes of the ImageNet dataset. The network used is $7$ layers deep with $49$ neurons per layer. The network is trained for $5000$ epochs. The images are resized to $256 \times 256$ pixels. Figure \ref{fig:imagenet} shows the rate-distortion curve for the ImageNet dataset. The bit rate is changed by changing the number of images to reconstruct. The lowest bit rate in Fig. \ref{fig:imagenet} is $0.025$ BPP, which is achieved by reconstructing $96$ images from $2$ sets of weights with an average PSNR of $18.79$ dB.

\subsection{Comparison with Existing Methods}
\begin{figure}[htbp]
    \centering
    \includegraphics[scale=0.4]{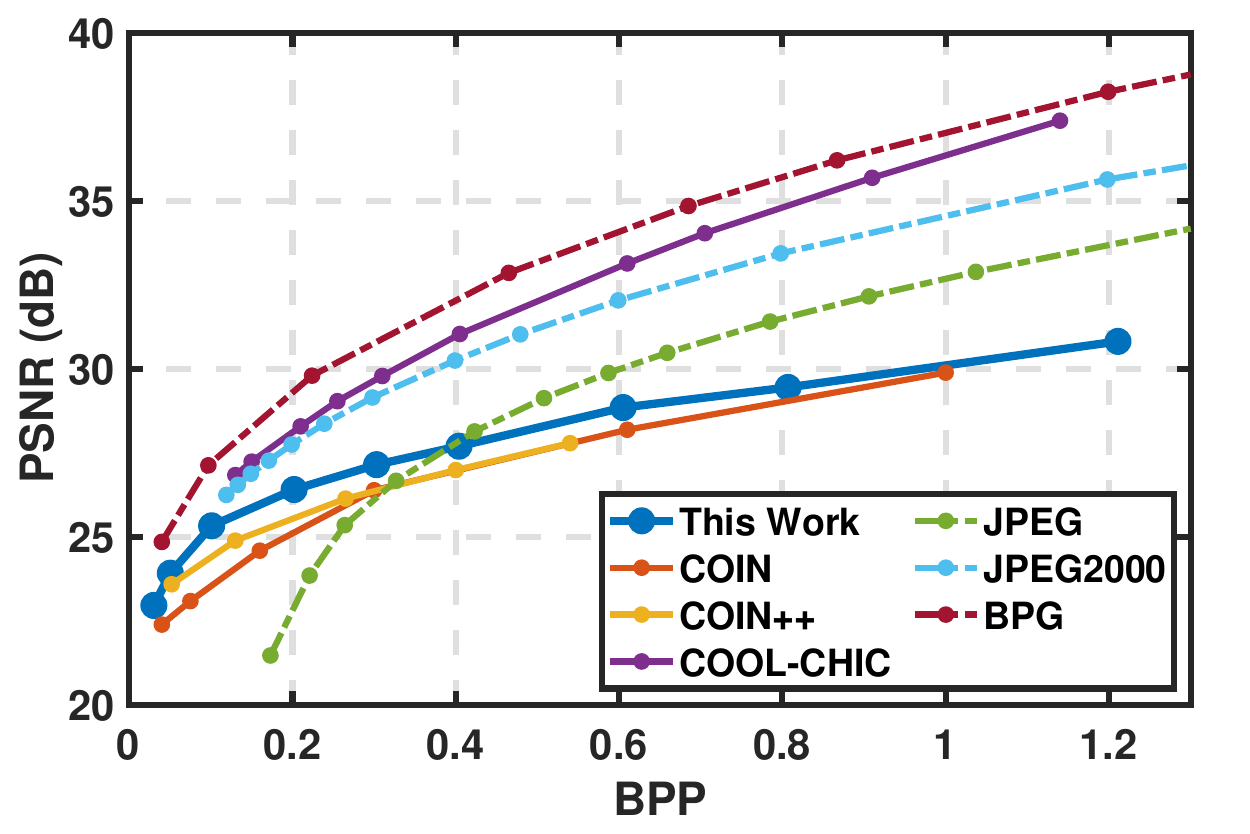} \vspace{-1em}
    \caption{Comparison of the rate-distortion curve of the proposed method with existing methods for the Kodak dataset.} \vspace{-0.5em}
    \label{fig:kodak_comparison}
\end{figure}

Fig. \ref{fig:kodak_comparison} compares the rate-distortion curve of the proposed method for the Kodak dataset with existing INR-based methods and standard image codecs, such as JPEG \cite{JPEG}, JPEG2000 \cite{JPEG2000}, and BPG \cite{BPG}. It is important to note that the reported bit rate of the proposed work does not include any entropy coding. While entropy coding may further reduce the bit rate, it is left for future work. The proposed method achieves a PSNR of over $25$ dB at a bit rate of $0.1$ BPP, which is comparable to existing methods. In fact, the proposed method achieves a lower bit rate than existing INR-based methods such as COIN \cite{Dupont2021} and COIN++ \cite{Dupont2022} while maintaining the same PSNR. INR-based methods along with entropy coding, such as COOL-CHIC \cite{Ladune2023}, achieve a higher PSNR at the same bit rate. However, the difference becomes smaller at lower bit rates. Therefore, the proposed method performs well compared to existing methods at low bit rates, \textit{i.e.}, when trying to generate a large number of images with a small number of weights.

\begin{figure}[htbp]
    \centering
    \includegraphics[scale=0.4]{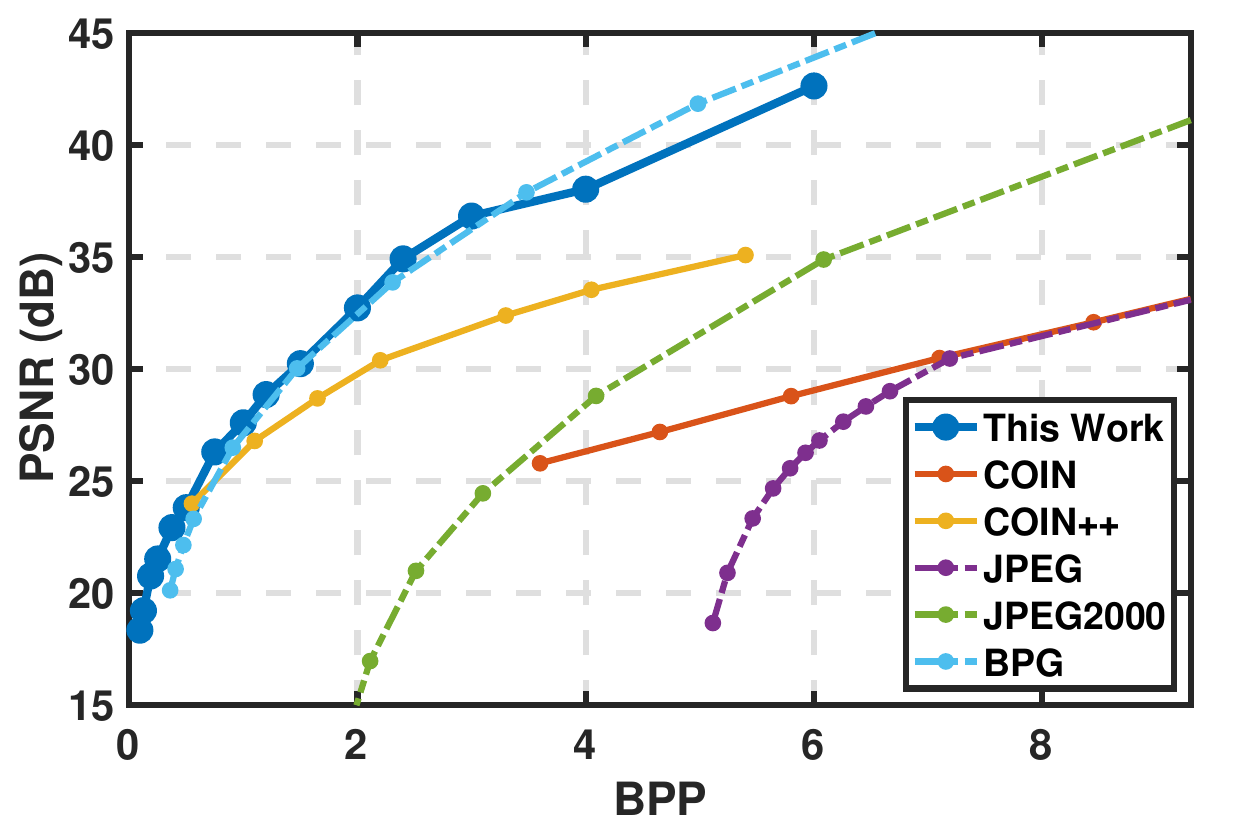} \vspace{-1em}
    \caption{Comparison of the rate-distortion curve of the proposed method with existing methods for the CIFAR-10 dataset.} \vspace{-1em}
    \label{fig:cifar10_comparison}
\end{figure}
Fig. \ref{fig:cifar10_comparison} shows the rate-distortion curve for the CIFAR-10 dataset compared with COIN, COIN++, and standard image codecs. As shown in Fig. \ref{fig:cifar10_comparison}, the proposed method outperforms INR-based methods and is comparable to state-of-the-art image codecs such as BPG. This is primarily due to the fact that entropy coding does not provide significant benefits for small images such as CIFAR-10. This observation implies that the proposed method, when combined with entropy coding, can close the gap between INR and standard codecs. However, further analysis is left for future work. \vspace{-1em}

\begin{figure*}[ht]
    \centering
    \includegraphics[scale=0.88]{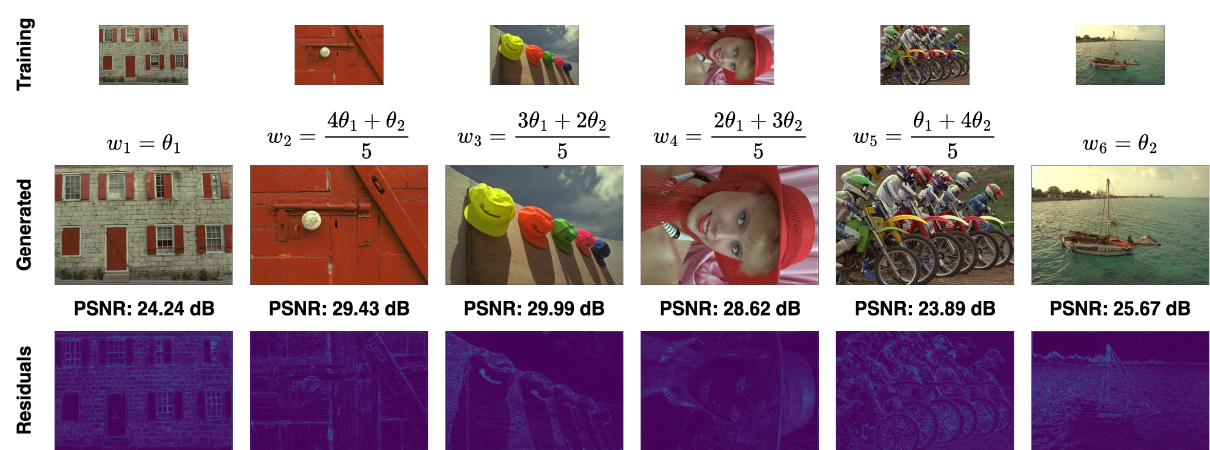}
    \caption{Downsampled images used for training, the generated images of the original size, and the residuals of the generated images of the first six images of the Kodak dataset. A network with $4$ hidden layers and $256$ neurons per layer is used, resulting in a BPP of $0.9$.}
    \label{fig:kodak_highres}
\end{figure*}
\subsection{Reconstruction of High-Resolution Images}
One key difference between INR-based image compression and algorithmic compression methods is that INR can reconstruct images at arbitrary resolutions. For instance, while the INR network weights can be trained for a $256 \times 256$ image, the step size of inputs during inference can be adjusted to reconstruct a $1024 \times 1024$ image. To validate this capability, the initial network with $4$ hidden layers and $256$ neurons per layer is trained on the first six images of the Kodak dataset, similar to the setup for Fig. \ref{fig:kodak_first_6}. During training, however, the images are downsampled by a factor of 2 in each dimension. During inference, the images are reconstructed at their original size to measure the PSNR. The results confirm that this capability is feasible, though it does come with a slight reduction in PSNR. The average PSNR is 26.97 dB. However, the training time is $4\times$ faster as the training images are $4\times$ smaller, leading to a trade-off between PSNR and training time. The BPP remains at $0.9$ as the network size and the generated image size are the same as in Fig. \ref{fig:kodak_first_6}.

Fig. \ref{fig:kodak_highres} shows the generated images and the residuals of the six images. The largest decrease in PSNR occurs in the fifth and sixth images, where there are numerous high-frequency or small details that are not clearly visible in the downsampled image. The same phenomenon would occur in regular INR, confirming that the core properties of INR remain intact with the proposed training method.
\section{Conclusion and Future Work}
\label{sec:conclusion}
This work proposed a novel method for training INRs using multiple sets of weights, addressing the fundamental limitation of one image per network. The validity of the proposed training method was theoretically established by proving the convergence of the loss functions. The proposed method was validated on various image datasets, such as Kodak, ImageNet, and CIFAR-10, achieving a PSNR of over $25$ dB at a bit rate of $0.1$ BPP, \textit{i.e.,} a compression factor of $80\times$, on the Kodak dataset. The proposed method was compared with existing INR-based methods and standard image codecs, showing that the proposed method performs well at low bit rates, and closes the gap with standard image codecs, outperforming prior INR-based methods for the CIFAR-10 dataset. It was shown that the entire Kodak dataset can be represented with just two sets of weights ($N=2$), and up to $128$ images from the CIFAR-10 dataset can be represented with two sets of weights. It was also shown that the properties of INR, such as the ability to reconstruct arbitrary-sized images during inference, are still maintained when using the proposed method. All the results in this work were obtained with $N=2$ and without any entropy coding. Future work includes increasing $N$ and $M$ (number of target images), incorporating entropy coding to further reduce the bit rate, and extending the proposed method to other applications. Another interesting avenue for future research is applying the proposed method in a decentralized setting, such as federated learning \cite{Savazzi2020, Guo2021, Nguyen2021, Ranjha2021, You2023}.


\begin{thebibliography}{00}

\bibitem{Wang2021} M. Wang, D. Xiao, and Y. Xiang, ``Low-Cost and Confidentiality-Preserving Multi-Image Compressed Acquisition and Separate Reconstruction for Internet of Multimedia Things,'' \textit{IEEE Internet of Things Journal}, vol. 8, no. 3, pp. 1662-1673, 2021.

\bibitem{Chakraborty2021} P. Chakraborty, J. Cruz, and S. Bhunia, ``MAGIC: Machine-Learning-Guided Image Compression for Vision Applications in Internet of Things,'' \textit{IEEE Internet of Things Journal}, vol. 8, no. 9, pp. 7303-7315, 2021.

\bibitem{Wang2022} Z. Wang, F. Li, J. Xu, and P. C. Cosman, ``Human–Machine Interaction-Oriented Image Coding for Resource-Constrained Visual Monitoring in IoT,'' \textit{IEEE Internet of Things Journal}, vol. 9, no. 17, pp. 16181-16195, 2022.

\bibitem{Chamain2022} L. D. Chamain, S. Qi, and Z. Ding, ``End-to-End Image Classification and Compression With Variational Autoencoders,'' \textit{IEEE Internet of Things Journal}, vol. 9, no. 21, pp. 21916-21931, 2022.

\bibitem{An2024} W. An, Z. Bao, H. Liang, C. Dong, and X. Xu, ``A Relay System for Semantic Image Transmission Based on Shared Feature Extraction and Hyperprior Entropy Compression,'' \textit{IEEE Internet of Things Journal}, vol. 11, no. 9, pp. 16158-16170, 2024.

\bibitem{Park2019} J. J. Park, P. Florence, J. Straub, R. Newcombe, and S. Lovegrove, ``DeepSDF: Learning Continuous Signed Distance Functions for Shape Representation,'' \textit{Proceedings of the IEEE Conference on Computer Vision and Pattern Recognition (CVPR)}, pp. 165-174, 2019.

\bibitem{Chen2019} Z. Chen and H. Zhang, ``Learning Implicit Fields for Generative Shape Modeling,'' \textit{Proceedings of the IEEE Conference on Computer Vision and Pattern Recognition (CVPR)}, pp. 5939-5948, 2019.

\bibitem{Mescheder2019} L. Mescheder, M. Oechsle, M. Niemeyer, S. Nowozin, and A. Geiger, ``Occupancy Networks: Learning 3D Reconstruction in Function Space,'' \textit{Proceedings of the IEEE Conference on Computer Vision and Pattern Recognition (CVPR)}, pp. 4460-4470, 2019.

\bibitem{Sitzmann2020} V. Sitzmann, J. N. P. Martel, A. W. Bergman, D. B. Lindell, and G. Wetzstein, ``Implicit Neural Representations with Periodic Activation Functions,'' \textit{Advances in Neural Information Processing Systems (NeurIPS)}, vol. 33, pp. 7462-7473, 2020.

\bibitem{Michalkiewicz2019} M. Michalkiewicz, J. K. Pontes, D. Jack, M. Baktashmotlagh, and A. Eriksson, ``Implicit Surface Representations As Layers in Neural Networks,'' \textit{International Conference on Computer Vision (ICCV)}, pp. 4743-4752, 2019.

\bibitem{Morreale2022} L. Morreale, N. Aigerman, P. Guerrero, V. G. Kim, and N. J. Mitra, ``Neural Convolutional Surfaces,'' \textit{Proceedings of the IEEE Conference on Computer Vision and Pattern Recognition (CVPR)}, pp. 19333-19342, 2022.

\bibitem{Chen2021} Y. Chen, S. Liu, and X. Wang, ``Learning Continuous Image Representation with Local Implicit Image Function,'' \textit{Proceedings of the IEEE Conference on Computer Vision and Pattern Recognition (CVPR)}, pp. 8624-8634, 2021.

\bibitem{Chen2022} Z. Chen et al., ``VideoINR: Learning Video Implicit Neural Representation for Continuous Space-Time Super-Resolution,'' \textit{Proceedings of the IEEE Conference on Computer Vision and Pattern Recognition (CVPR)}, pp. 2037-2047, 2022.

\bibitem{Luigi2023} L. D. Luigi, A. Cardace, R. Spezialetti, P. Z. Ramirez, S. Salti, and L. D. Stefano, ``Deep Learning on Implicit Neural Representations of Shapes,'' \textit{International Conference on Learning Representations (ICLR)}, 2023.

\bibitem{Mildenhall2020} B. Mildenhall, P. P. Srinivasan, M. Tancik, J. T. Barron, R. Ramamoorthi, and R. Ng, ``NeRF: Representing Scenes as Neural Radiance Fields for View Synthesis,'' \textit{European Conference on Computer Vision (ECCV)}, pp. 405-421, 2020.

\bibitem{Hornik1989} K. Hornik, M. Stinchcombe, and H. White, ``Multilayer feedforward networks are universal approximators,'' \textit{Neural Networks}, vol. 2, no. 5, pp. 359-366, 1989.

\bibitem{Cybenko1989} G. Cybenko, ``Approximation by superpositions of a sigmoidal function,'' \textit{Mathematics of Control, Signals, and Systems}, vol. 2, no. 4, pp. 303-314, 1989.

\bibitem{Dupont2021} E. Dupont, A. Goliński, M. Alizadeh, Y. W. Teh, and A. Doucet, ``COIN: COmpression with Implicit Neural representations,'' \textit{International Conference on Learning Representations (ICLR)}, 2021.

\bibitem{Dupont2022} E. Dupont, H. Loya, M. Alizadeh, A. Goliński, Y. W. Teh, and A. Doucet, ``COIN++: Neural Compression Across Modalities,'' \textit{Transactions on Machine Learning Research}, vol. 2022, no. 11, 2022.

\bibitem{Strumpler2022} Y. Str\"umpler, J. Postels, R. Yang, L. V. Gool, and F. Tombari, ``Implicit Neural Representations for Image Compression,'' \textit{European Conference on Computer Vision (ECCV)}, pp. 74-91, 2022.

\bibitem{Ladune2023} T. Ladune, P. Philippe, F. Henry, G. Clare, and T. Leguay, ``COOL-CHIC: Coordinate-based Low Complexity Hierarchical Image Codec,'' \textit{International Conference on Computer Vision (ICCV)}, pp. 13515-13522, 2023.

\bibitem{Pham2023} T. Pham, Y. Yang, and S. Mandt, ``Autoencoding Implicit Neural Representations for Image Compression,'' \textit{International Conference on Machine Learning (ICML)}, 2023.

\bibitem{Lee2023} S. Lee, J. -B. Jeong, and E. -S. Ryu, ``Entropy-Constrained Implicit Neural Representations for Deep Image Compression,'' \textit{IEEE Signal Processing Letters}, vol. 30, pp. 663-667, 2023.

\bibitem{Girish2023} S. Girish, A. Shrivastava, and K. Gupta, ``SHACIRA: Scalable HAsh-grid Compression for Implicit Neural Representations,'' \textit{International Conference Computer Vision (ICCV)}, pp. 17513-17524, 2023.



\bibitem{TranFL} N. H. Tran, W. Bao, A. Zomaya, M. N. H. Nguyen, and C. S. Hong, ``Federated learning over wireless networks: Optimization model design and analysis,'' \textit{Proc. IEEE Int. Conf. Comput. Commun. (INFOCOM)}, pp. 1387–1395, 2019.

\bibitem{AdaptFL} S. Wang et al., ``Adaptive federated learning in resource constrained edge computing systems,'' \textit{IEEE J. Sel. Areas Commun. (JSAC)}, vol. 37, no. 6, pp. 1205–1221, 2019.

\bibitem{DinhFL} C. Dinh et al., ``Federated learning over wireless networks: Convergence analysis and resource allocation,'' \textit{IEEE/ACM Transactions on Networking}, vol. 29, no. 1, pp. 398-409, Feb. 2021.

\bibitem{Hybrid} M. Friedlander and M. Schmidt, ``Hybrid deterministic-stochastic methods for data fitting,'' \textit{SIAM Journal on Scientific Computing}, vol. 34, no. 3, pp.~A1380-A1405, 2012.

\bibitem{Kodak} E. Kodak, ``Kodak Lossless True Color Image Suite,'' \url{https://r0k.us/graphics/kodak/} (accessed Apr. 18, 2024).

\bibitem{ImageNet} J. Deng, W. Dong, R. Socher, L. -J. Li, K. Li, and L. Fei-Fei, ``ImageNet: A large-scale hierarchical image database,'' \textit{IEEE Conference on Computer Vision and Pattern Recognition}, pp. 248-255, 2009.

\bibitem{CIFAR10} A. Krizhevsky, ``Learning Multiple Layers of Features from Tiny Images,'' \textit{Technical Report, University of Toronto}, 2009.

\bibitem{JPEG} G. K. Wallace, ``The JPEG still picture compression standard,'' \textit{IEEE Transactions on Consumer Electronics}, vol. 38, no. 1, pp. xviii-xxxiv, Feb. 1992.

\bibitem{JPEG2000} A. Skodras, C. Christopoulos, and T. Ebrahimi, ``The JPEG 2000 still image compression standard,'' \textit{IEEE Signal Processing Magazine}, vol. 18, no. 5, pp. 36-58, Sept. 2001.

\bibitem{BPG} F. Bellard, ``BPG Image format,'' \url{https://bellard.org/bpg/} (accessed May 4, 2024).

\bibitem{Savazzi2020} S. Savazzi, M. Nicoli, and V. Rampa, ``Federated Learning With Cooperating Devices: A Consensus Approach for Massive IoT Networks,'' \textit{IEEE Internet of Things Journal}, vol. 7, no. 5, pp. 4641-4654, 2020.

\bibitem{Guo2021} H. Guo, A. Liu, and V. K. N. Lau, ``Analog Gradient Aggregation for Federated Learning Over Wireless Networks: Customized Design and Convergence Analysis,'' \textit{IEEE Internet of Things Journal}, vol. 8, no. 1, pp. 197-210, 2021.

\bibitem{Nguyen2021} V. -D. Nguyen, S. K. Sharma, T. X. Vu, S. Chatzinotas, and B. Ottersten, ``Efficient Federated Learning Algorithm for Resource Allocation in Wireless IoT Networks,'' \textit{IEEE Internet of Things Journal}, vol. 8, no. 5, pp. 3394-3409, 2021.

\bibitem{Ranjha2021} A. Ranjha and G. Kaddoum, ``URLLC Facilitated by Mobile UAV Relay and RIS: A Joint Design of Passive Beamforming, Blocklength, and UAV Positioning,'' \textit{IEEE Internet of Things Journal}, vol. 8, no. 6, pp. 4618-4627, 2021.

\bibitem{You2023} C. You, K. Guo, G. Feng, P. Yang, and T. Q. S. Quek, ``Automated Federated Learning in Mobile-Edge Networks—Fast Adaptation and Convergence,'' \textit{IEEE Internet of Things Journal}, vol. 10, no. 15, pp. 13571-13586, 2023.

\end{thebibliography}
\end{document}